\documentclass[aps]{revtex4}
\usepackage{epsfig}
\begin{document}
\preprint{USACH-FM-02-04}
\title{ Testing  spatial noncommutativity via the Aharonov-Bohm effect}
\author{ H. Falomir$^1$\thanks{E-mail: falomir@obelix.fisica.unlp.edu.ar},
J. Gamboa$^{2}$\thanks{E-mail:
jgamboa@lauca.usach.cl}, M. Loewe$^{3}$\thanks{E-mail:
mloewe@fis.puc.cl,}, F. M\'endez$^{2}$\thanks{E-
mail:fmendez@lauca.usach.cl} and J. C. Rojas$^2$\thanks{E-mail:
jrojas@lauca.usach.cl}}
\address{$^1$ IFLP - Departamento de F\'{\i}sica, Facultad de Ciencias Exactas,
Universidad Nacional de la Plata,
La Plata, Argentina\\
$^2$Departamento de F\'{\i}sica, Universidad de Santiago de Chile,
Casilla 307, Santiago 2, Chile\\
$^3$ Facultad  de F\'{\i}sica, Pontificia Universidad  Cat\'olica de
Chile, Casilla 306, Santiago 22, Chile}
\begin{abstract}
The possibility of detecting noncommutative space relics is
analyzed using the Aharonov-Bohm effect. We show that, if space is
noncommutative, the holonomy receives non-trivial kinematical
corrections that will produce a diffraction pattern even when the
magnetic flux is quantized. The scattering problem is also
formulated, and the differential cross section is calculated. Our
results can be extrapolated to high energy physics and the bound
$\theta \sim [ 10 \,\mbox{TeV}]^{-2}$ is found. If this bound
holds, then noncommutative effects could be explored in scattering
experiments measuring differential cross sections for small
angles. The bound state Aharonov- Bohm effect is also discussed.
\end{abstract}
\maketitle
\section{Introduction}
There are arguments in string theory suggesting that  spacetime
could be noncommutative \cite{string}. Although this property
might be an argument in favor of new renormalizables effective
field theories \cite{various}, it represents also a trouble
because we need to explain the transition between the commutative
and noncommutative regimes.

If the noncommutative effects are important at very high energies,
then one could posit a decoupling  mechanism producing
the standard quantum field theory as an effective field theory having no memory about
noncommutative effects. However, our experience in atomic and
molecular physics \cite{mol} strongly suggests that the decoupling
is never complete, and the high energy effects appear in the
effective action as topological remnants \cite{fw}.

Following this idea we would like to consider an example, related
to topological aspects, where the appearance of noncommutative
effects could be relevant. A natural candidate is the
Aharonov-Bohm effect \cite{ab} where, as we know, the relativistic
corrections do not change the qualitative behavior of the fringe
pattern \cite{vitorio}.

    As we will see,
if the space is noncommutative the total holonomy contains  --as we will show below--
a term dependent on the velocity of the electrons, which tends to
shift the line spectrum.  Moreover, a new effect is produced by noncommutativity: Particles are scattered
even when the magnetic flux is quantized.

Our conclusions are reinforced by studying the bound state
Aharonov-Bohm effect \cite{pesh}. In this case, although the
Schr\"odinger equation cannot be exactly solved, one can extract information
through perturbation theory since $\theta<<1$.

As a bonus of the previous results,  one find -- using
perturbation theory--  an explicit expression for the scattering
amplitude and a  formula for the differential cross section of the
noncommutative Aharonov-Bohm effect.

There is, however, another interesting conclusion that can be
extracted from our research.  The quantum mechanical Aharonov-Bohm
effect is also a relevant mechanism to explain other high energy
phenomena. In this sense, our calculations allow to extract
conclusions for other high energies processes, {\it e.g.}   cosmic
strings and GUTs \cite{aw}. More precisely, using our
noncommutative differential cross section, we are able to find a
bound for the theta parameter which is in full agreement with 
other estimations \cite{carroll}.

The paper is organized as follows: in section 2, the
noncommutative Aharonov-Bohm effect is discussed  and a formula
for the holonomy is derived; in section 3 we explain the
noncommutative corrections to the bound state Aharonov- Bohm
effect.  The general Schr\"odinger equation and the scattering
problem in a noncommutative space are considered in section 4;  in
section 5, we studied the first order noncommutative corrections
to the scattering amplitude; in section 6,   we estimate a bound
for the noncommutative parameter and we analyse the experimental
possibilities for detecting noncommutative relics  and, finally,
section 7 contains the conclusions. Two appendixes containing a
discussion on the commutative Aharonov-Bohm effect at high energy
and some technical details are included.

\section{The Noncommutative Aharonov-Bohm Effect}

\subsection{The Schr\"odinger equation}

    In the commutative case, the Schr\"odinger equation with an external gauge potential
is solved by
\begin{equation}
\psi = e^{i\int_C dx^j A_j} \varphi, \label{9}
\end{equation}
where $\varphi$ is the solution of the free Schr\"odinger
equation, and the $U(1)$ holonomy, $ e^{i\int_C dx^j A_j}$, is in
general a non-integrable factor, {\it i.e.} it depends on the
integration path $C$.

Although eq.\ (\ref{9}) solves formally the Schr\"odinger
equation, the holonomy  involves in a non trivial way the dynamics
of the gauge potential, hiding all the complications related to
${\bf A}$. Our goal below will be to find an approximate
expression for the holonomy, valid for small values of the
fundamental noncommutative parameter $\theta$.

    In the following we assume that wavefunctions in the plane belong to a
    noncommutative algebra characterized by the Moyal product,  defined as
     \begin{equation}
     \left({\bf A}\star {\bf B}\right)({\bf x}) = e^{\frac{i}{2}\theta\, \epsilon^{ij}
     \partial^{(1)}_i \partial^{(2)}_j}
     {\bf A}({\bf x}_1) {\bf B}({\bf x}_2) \vert_{{\bf x}_1={\bf
      x}_2=
     {\bf x}} \label{moyal}.
    \end{equation}

The  Schr\"odinger equation in this noncommutative space is
\begin{equation}
{\hat H} \psi = \frac{1}{2m} D_j \star D_j \star \psi =
\frac{1}{2m}k_j k_j \, \psi. \label{11}
\end{equation}
where $k_j $ are the eigenvalues of the operator  $D_j = -i
\partial_j + A_j$, {\it i.e.}
\begin{equation}
D_j \star \psi = k_j \, \psi, \label{10}
\end{equation}
we are assuming, of course, that the magnetic field is zero everywhere except in the origin. 

In order to solve (\ref{10}) we use the Ansatz
\begin{equation}
\psi = e^{F}. \label{12}
\end{equation}
As we are assuming that  noncommutative effects are small, we
expand the Moyal product retaining only  linear terms in $\theta$,
\begin{eqnarray}
D_j \star \psi&=& -i\partial_j\, e^F +  A_j \star e^F \nonumber
\\
&= & e^F [ -i \partial_j F +  A_j + \frac{i}{2}  \theta
\epsilon^{lm} (\partial_l A_j)(\partial_m F) ] \nonumber.
\end{eqnarray}
Then (\ref{10}) becomes
\begin{equation}
-i \partial_j F + A_j + \frac{i}{2}  \theta \epsilon^{lm}
(\partial_l A_j) (\partial_m F) = k_j. \label{13}
    \end{equation}
Now, one can solve (\ref{13}) perturbatively expanding  $F$ and
$A_j$ in powers of $\theta$, {\it i.e.}
\begin{eqnarray}
F &=& F^{(0)} + \theta\, F^{(1)} + ... \label{14}
\\
A_j &=& A^{(0)}_j + \theta \, A_j^{(1)} + ....\label{144}
\end{eqnarray}

At zero order in $\theta$, equation (\ref{13}) gives
\begin{equation}
-i \partial_j F^{(0)} +  A^{(0)}_j = k_j, \label{15}
\end{equation}
from which the following expression for $F^{(0)}$ is obtained:
\begin{equation}
 F^{(0)} = ik_j (x-x_0)_j - i \int_{x_0}^x dx_j A^{(0)}_j. \label{16}
 \end{equation}

The first term in the RHS is just the free particle solution if we
interpret $k_j$ as the wave number, and the second term is the
$U(1)$ holonomy for the commutative case. Thus, at zero order we
reproduce the solution of the
    commutative case
Schr\"odinger equation.

If we retain first order terms in $\theta$, the following
differential equation is obtained,
\begin{equation}
-i\partial_j F^{(1)} + A^{(1)}_j + \frac{i}{2}  \epsilon^{lm}
(\partial_l  A^{(0)}_j) (\partial_m F^{(0)} )= 0, \label{17}
\end{equation}
which by integration gives
\begin{equation}
 F^{(1)} = - i \int_{x_0}^x dx_j A^{(1)}_j - \frac{i}{2}
 \int_{x_0}^x dx_j\, \epsilon^{ml} k_m \,
\partial_l A^{(0)}_j + \frac{i}{2}
 \int_{x_0}^x dx_j\, \epsilon^{ml}  A^{(0)}_m \,
\partial_l A^{(0)}_j. \label{18}
 \end{equation}

The first term in the RHS of (\ref{18}) is an additive correction
to the commutative holonomy which, together with the second term
in the RHS of (\ref{16}) gives
\begin{equation} - i \int_{x_0}^x
dx_j ( A^{(0)}_j + \theta A^{(1)}_j). \nonumber
\end{equation}
The second in the RHS of (\ref{18}) is a velocity dependent term,
which can be written as  \footnote{A different expression was
obtained in \cite{chaichian}, although later it was corrected by
these authors who agree with our previous result \cite{glr}.}
\begin{equation}\label{segundo}
- \frac{i}{2}  \int_{x_0}^x dx_j \,\,\epsilon^{ml} k_m \partial_l
A^{(0)}_j = -\frac{i}{2}m \int_{x_0}^x \,dx_j \, ({\bf v} \times
\nabla A^{(0)}_j)_3. \label{19}
\end{equation}
For the last term our calculation yields
\begin{equation}\label{tercero}
     \frac{i}{2}
    \int_{x_0}^x dx_j\, \epsilon^{ml}  A^{(0)}_m \,
    \partial_l A^{(0)}_j = \frac{i}{2}\int_{x_0}^x dx_j \,
    ({\bf A}^{(0)}\times \nabla
    A^{(0)}_j)_3.\nonumber
\end{equation}

Thus,  at this order in $\theta$, the nonconmutative holonomy is
given by
\begin{equation}
    {\cal W}(x,x_0) =\exp\left\{-i\biggl[ \int_{x_0}^x dx_j A_j
    + \frac{\theta}{2}\,  \int_{x_0}^x dx_j \,\left[ m \, ({\bf v}\times
    \nabla A^{(0)}_j)_3  - ({\bf A^{(0)}}\times \nabla A^{(0)}_j)_3
    \right]\biggr]\right\}. \label{21}
\end{equation}

Now, we analyze the different terms in (\ref{21}): The first one
in the exponential is the usual holonomy, corrected to order
$\theta$, which classifies the different  homotopy classes. The
term in (\ref{tercero}) is a noncommutative correction to the
vortex decaying as $1/r^3$, which does not contribute to the line
spectrum. Finally, the term in (\ref{segundo}) is a velocity
dependent correction insensitive to the topology of the manifold.

In the commutative Aharonov-Bohm effect, the presence of the flux
produces a shift in the interference pattern, which is maximum for
$\Phi=(2n+1)\pi (\hbar c/e)$, with $n \in Z$.  In such case, for a
given value of  $n$,  the position of maxima and minima are
interchanged due  to a change of $\pi$ in the phase. However, in
the noncommutative case, this change of positions of maxima and
minima might not occur: Indeed, the velocity dependent correction
modifies the phase shift which, for suitable values of velocity,
could even become $2\pi$ for a given $n$.

\vskip 1cm

We finalize this section emphasizing two importants aspects of our
results:
\begin{itemize}
\item{} The above results are a general property of the noncommutative
Aharonov-Bohm effect, depending only on the total flux $\Phi$ (if the electrons can not penetrate into the solenoid).
\item{} If the magnetic flux $e\Phi/h c$ is an
integer there is no Aharonov-Bohm effect for the commutative case,
as is well known \cite{ab,rui}. However, in the noncommutative
case the term (\ref{segundo}) is different from zero
 {\bf even in the case where $e \Phi/hc $ is an integer}. This a
quite non-trivial characteristic of the noncommutative
Aharonov-Bohm effect that could be experimentally measured.
\end{itemize}

\subsection{The Gauge Potentials}

In this section we will evaluate the gauge potential for a finite
radius solenoid orthogonal to a noncommutative plane.

The field tensor in the noncommutative plane is
\begin{equation}
{\hat F}_{\mu \nu} = \partial_\mu A_\nu - \partial_\nu A_\mu + i
A_\mu \star A_\nu - i  A_\nu \star A_\mu. \label{1}
\end{equation}

Expanding the Moyal product and retaining only the linear term in
$\theta$, we have
\begin{equation}
{\hat F}_{\mu \nu} = \partial_\mu A_\nu - \partial_\nu A_\mu +
 \theta \, \epsilon^{\alpha \beta} \partial_\alpha A_\mu \partial_\beta
A_\nu. \label{2}
\end{equation}

We must construct a gauge potential such that the magnetic field
$B_3={\hat F}_{1 2}$ vanishes everywhere, except inside the
solenoid. We proceed as in the commutative case, starting with the
Ansatz (where the ordinary product is understood)
\begin{eqnarray}
A_1 &=& - x_2 f(r^2), \nonumber
\\
A_2 &=& x_1 f(r^2), \label{4}
\end{eqnarray}
for $r>a$, the radius of the solenoid.

    We impose the condition  $B_3 = {\hat F}_{12}=0$ outside
the solenoid, implying that
\begin{equation}
 2 f + 2 r^2 f^{'} +  \theta ( f^2 + 2 r^2 f f^{'}) = 0, \label{5}
\end{equation}
where $f^{'} = df/dr^2$.

This differential equation can be easily integrated, yielding  the
following solutions
\begin{eqnarray}
f &=& - \frac{1}{  \theta} \pm \frac{1}{  \theta} \sqrt{1 +
\frac{c_1\theta}{r^2}} \nonumber
\\
&=&- \frac{1}{ \theta} \pm  \frac{1}{  \theta} [ 1 +
\frac{c_1\theta}{2r^2} - \frac{ c_1^2 \theta^2}{8r^4} + ...],
\label{6}
\end{eqnarray}
where $c_1$ is an integration constant.

From (\ref{6}) we see that the commutative limit is smooth for the
plus sign in the above equation, then we adopt
\begin{equation}
 f = \frac{c_1}{2r^2} - \frac{c^2_1 \theta}{8r^4} + ...\label{7}
\end{equation}

We determine the integration constant by imposing the Stokes
theorem at zero order in $\theta$,
\begin{equation}\label{stokes}
  \int\!\!\!\!\!\int {\vec B}\cdot d{\vec
S}=\Phi = \oint {\vec A^{(0)}}\cdot d{\vec l} ,
\end{equation}
getting
\[
c_1 = \frac{\Phi}{\pi} = B a^2.
\]
Notice that eq.\ (\ref{6}) requires $B\, \theta \ll 1$.

The final expression for the gauge potential becomes
\begin{eqnarray}
A_1 &=&  x_2 \biggl( \frac{\Phi}{2\pi(x_1^2 + x_2^2)} -\theta\,
\frac{
 \Phi^2}{8\pi^2(x_1^2 + x_2^2)^2} + ...\biggr),\nonumber
\\
A_2 &=&  -x_1 \biggl( \frac{\Phi}{2\pi(x_1^2 + x_2^2)} -\theta\,
\frac{
 \Phi^2}{8\pi^2(x_1^2 + x_2^2)^2} + ...\biggr), \label{8}
\end{eqnarray}
where $\Phi=B\pi a^2$ is the magnetic flux enclosed into the
solenoid.

We finally give the gauge potential expressed in terms of polar
coordinates,
\begin{eqnarray}
A_r &=& 0,\nonumber
\\
A_{\varphi}&=& \frac{\Phi}{2\pi r} - \theta\,\frac{ \Phi^2}{8\pi^2
r^3} + {\cal O}(\theta^2), \label{aphi}
\end{eqnarray}
which will be useful in solving the Schr\"odinger equation in the
next sections.

\section{ Bound  States for  the  Noncommutative
Aharonov-Bohm Effect}

In this section we will solve the noncommutative Schr\"odinger
equation (\ref{11}) for an electron moving in a two-dimensional
manifold  parametrized by polar coordinates $(r,\varphi)$, with $r
> a$ and $0 < \varphi < 2\pi $.

Before doing this, it is necessary to explain an important
technical point: The Moyal product (\ref{moyal}) is implicitly
written in cartesian coordinates. Therefore,  in order to solve
(\ref{11}) in polar coordinates, one must express the $\star$
product in the general case.

We find the following expression for the  the Moyal product  up to
first order in $\theta$
\begin{equation}
f(x) \star p(x) = f(x)\,  p(x) + \frac{i\theta}{2\sqrt{g}} \,
\epsilon^{\mu \nu} \, \partial_\mu f \, \partial_\nu p + {\cal
O}(\theta^2), \label{mod}
\end{equation}
where $g$ is the determinant of the metric.

At this order in $\theta$, the Schr\"odinger equation becomes
\begin{equation}
(\hat{H_0} +\theta\,\hat{H_1})\Psi  =k^2 \Psi, \label{schteta}
\end{equation}
where $H_0$ and  $H_1$
    can be identified by replacing (\ref{aphi}) in (\ref{11}), and
    taking into account that the covariant derivative becomes
    \begin{equation}
    {\vec D} = -i \left[ {\hat r}\, \partial_r + {\hat \varphi}\,
    \left(\frac{\partial_\varphi}{r} - i A_\varphi \right)\right].
    \nonumber
    \end{equation}
We get
\begin{eqnarray}
\hat{H_0}&=& \partial^2 _r + \frac{1}{r}\, \partial_r +
\frac{1}{r^2}\left(\,\partial^2_\varphi - i\frac{\Phi}{\pi}
\partial _\varphi - \frac{\Phi^2}{4\pi^2}
\right), \label{h0} \\
\hat{H_1}&=&\frac{1}{r^3}\left(-i\partial _\varphi -
\frac{\Phi}{4\pi}\right)\partial _r + \frac{1}{r^4}\left( -i
\partial^3_\varphi-\frac{\Phi}{\pi}\partial^2_\varphi +
i\frac{\Phi^2}{2\pi^2}
\partial _\varphi +\frac{\Phi^3}{8\pi^3}\right).  \label{h1}
\end{eqnarray}

As $\theta$ is very small, one can use perturbation theory for
computing the eigenvalues and eigenfunctions of the hamiltonian
$ \hat{H}=\hat{H_0} + \theta \, \hat{H_1} $.

In the following subsections, we find explicitly the energy
spectrum for the bound state Aharonov-Bohm effect.

\subsection{The  Noncommutative  Bound State Aharonov-Bohm Effect}

The bound state  Aharonov-Bohm effect is a result due to Peshkin
{\it et. al.} \cite{pesh}, which establishes the flux
    and angular momentum
dependence of the energy spectrum, a measurable quantity in principle.
In this effect one consider an electron constrained to move
between  two impenetrable concentric cylinders with outer and
inner radius $b$ and $a$ respectively,
    and in the presence of a magnetic flux $\Phi$ contained inside the inner one.

In the noncommutative space, the Schr\"odinger radial equation  at
first order in $\theta$ is given by
\begin{equation}
\begin{array}{c}
    \displaystyle{
    -\chi_{\ell}''(r) -  \left( \frac{1}{r} + \frac{\theta \,\left(
    \ell - \frac{ \Phi }{4\,\pi } \right) }{r^3} \right) \,\chi_\ell'(r)
    + \left( \frac{({\ell - \frac{\Phi}{2\pi})}^2}{r^2} +
    \frac{\theta \,\left( \ell^3 - \frac{\ell^2\,\Phi }{\pi } +
    \frac{\ell\,{\Phi }^2}{ 2\,{\pi } ^2} - \frac{{\Phi }^3}{8\,{\pi
    }^3}\right) }{r^4} \right) \,\chi_\ell (r) = }\\ \\ =
    \displaystyle{
    \left( \hat{h}_{\ell,0}+ \theta\, \hat{h}_{\ell,1} \right)
    \chi_{\ell}(r)
    = k^2 \, \chi_{\ell}(r)}, \label{fer2}
\end{array}
\end{equation}
where we have called
\begin{eqnarray}
\hat{h}_{\ell,0}&=& -\partial^2 _r - \frac{1}{r}\, \partial_r +
\frac{1}{r^2} \,\left(\ell - \frac{\Phi}{2 \pi}\right)^2,  \\
\hat{h}_{\ell,1}&=&-\frac{1}{r^3}\,\left(\ell -
\frac{\Phi}{4\pi}\right)\,\partial _r + \frac{1}{r^4}\,\left(
\ell^3 - \frac{\ell^2 \Phi}{ \pi} + \frac{\ell \Phi^2}{2 \pi^2}-
\frac{\Phi^3}{8 \pi^3} \right), \label{h11}
\end{eqnarray}
and the  following Ansatz has been used for the wave function:
\begin{equation}
\psi (r, \varphi) = \sum_{\ell \in Z } e^{i \ell \varphi}
\chi_\ell (r). \label{fer}
\end{equation}

Although this equation cannot be solved exactly,
    one can use perturbation theory in the small parameter
    $\theta$. Since noncommutative effects are important only at
    small distances $\sim \sqrt{\theta}$, one would expect some
    relevant consequences in the high energy region, $k\sim 1/\sqrt{\theta}$.

Equation (\ref{fer2}) contains  the commutative Aharonov-Bohm
effect as a particular case, for $\theta =0$. The 0-th order
solution can be written as
\begin{equation}
\chi_\ell (r) = A_\ell J_\nu (kr)  + B_\ell Y_\nu (kr),
\label{fer3}
\end{equation}
with $\nu= \vert \ell - \frac{\Phi}{2\pi}\vert$. The constants
$A_\ell$, $B_\ell$ and the Hamiltonian eigenvalues
$E_{\ell,0}=k^2$ can be obtained --as usual-- by imposing the
boundary conditions on $\chi_\ell(r)$,
\begin{equation}
\chi_\ell (a) =0= \chi_\ell (b), \label{fer33}
\end{equation}
together with the normalization condition for the eigenfunction,
\begin{equation}\label{norma1}
  \int_a^b \chi_\ell(r)^2 \, r\, dr = 1.
\end{equation}

    Notice that the eigenvalues depend on the angular momentum $\ell$
    only through $\nu$. Therefore, degeneracy will occur if $\vert
    \ell_1 - \frac{\Phi}{2\pi}\vert = \vert \ell_2 -
    \frac{\Phi}{2\pi}\vert$, which is possible only if $\Phi/\pi$ is
    an integer. For simplicity, to be able to apply perturbation
    theory in its simplest form, in this Section we will avoid these
    particular values of the flux.

    Taking into account (\ref{fer33}), the mean value of
    $\hat{h}_{\ell,1}$ can be straightforwardly cast in the form
    \begin{equation}\label{h1-val-med}
      E_{\ell,1}=\int_a^b \chi_\ell(r)\,\hat{h}_{\ell,1}
      \chi_\ell(r)\,r\,dr =
      P(\ell,\Phi) \,
     \langle r^{-4}\rangle_\ell,
    \end{equation}
    where $P(\ell,\Phi)$ is a cubic polynomial,
    \begin{equation}\label{poly}
    P(\ell,\Phi)=\left[\ell^3 - \frac{\ell^2\,\Phi }{\pi } + \ell \left(
     \frac{{\Phi }^2}{2\,{\pi }^2} -1  \right)  +
     \frac{2\,{\pi }^2\,\Phi  - {\Phi }^3}{8\,{\pi }^3}\right],
    \end{equation}
    and
    \begin{equation}\label{r-med}
       \langle r^{-4}\rangle_\ell =
     \int_a^b  \frac{\chi_\ell^2(r)}{r^4} \, r \,dr
    \end{equation}
    is a function of $\nu,a,b$ and $k$ only.

Since noncommutative effects are expected to occur at high
energies ($k\, a >> 1$), it is enough to use in (\ref{fer3}) the first terms in
the asymptotic expansions
    of Bessel functions for large arguments. We will retain just
the first two terms in these expansions, {\it i.e.}
\begin{eqnarray}
J_\nu (z) &\rightarrow& {\sqrt{\frac{2}{z\,\pi }}}\, \left[ \cos
\left( z - \frac{\pi \,\nu }{2}-\frac{\pi }{4}\right) -
\frac{4\nu^2-1}{8\, z} \sin\left(z - \frac{\pi \,\nu
}{2}-\frac{\pi }{4} \right)\right], \nonumber
\\
Y_\nu (z) &\rightarrow& -{\sqrt{\frac{2}{ z\,\pi }}}\, \left[ \sin
\left( z - \frac{\pi \,\nu }{2}-\frac{\pi }{4} \right)+
\frac{4\nu^2-1}{8\, z} \cos \left( z - \frac{\pi \,\nu
}{2}-\frac{\pi }{4}\right) \right]. \label{fer4}
\end{eqnarray}

Using  (\ref{fer4})
    in (\ref{norma1}) and (\ref{r-med}), we get
\begin{eqnarray}
\langle r^{-4}\rangle_\ell &=& \frac{128 a^3 k^5}
     {D(\nu,a,b,k)}  \times \nonumber
\\
&&\bigg \{ -\frac{1}{768 a^6 b^3 k^4}\bigg(512a^5k^2
+8a^3(1-4\nu^2)^2 +128 a^4b^3k^4(4\nu^2 -9) + b^3(1-4\nu^2)^2 (7
+4\nu^2) + \nonumber
\\
&&(2a^2b^3k^2)(116\nu^2 + 48\nu^4 -64\nu^6 -31)\bigg) -\nonumber
\\
&&\frac{\cos[2(a-b)k] }{768 a^3 b^4
k^4}\bigg(b(2b^2k^2-1)(1-4\nu^2)^2 -64a^2bk^2(4\nu^2-9)(2b^2k^2-1)
+ \nonumber
\\
&&8a(4\nu^2-1)(3-12\nu^2+2b^2k^2 (4\nu^2-9)) \bigg) +
\frac{\sin[2k(a-b)]}{1536 a^3 b^4 k^5}\bigg(  (1-4\nu^2)^2(3-12
\nu^2 +2b^2k^2(4\nu^2-9))  - \nonumber
\\
&&64a^2k^2(3-12\nu^2+2b^2k^2(4\nu^2-9) )
-32abk^2(-1+2b^2k^2)(9-40\nu^2+16\nu^4) \bigg) +\nonumber
\\
&&\mbox{Ci}(2ak) \bigg( \cos(2ka)\bigg[\frac{16\nu^4 -40\nu^2
+9}{12a^2} \bigg]+\sin(2ka)\bigg[\frac{(4\nu^2 -9) (64 a^2 k^2-(1
-4\nu^2)^2 )}{192 a^3 k}\bigg] \bigg)+ \nonumber
\\
&&  \mbox{Ci}(2bk) \bigg( \sin(2ka)\bigg[\frac{(4\nu^2 -9)(-64 a^2
k^2+(1 -4\nu^2)^2 )}{192 a^3 k}\bigg] -\cos(2ka)
\bigg[\frac{16\nu^4 -40\nu^2 +9}{12a^2} \bigg]\bigg)+ \nonumber
\\
&&\mbox{Si}(2ak) \bigg( \sin(2ka)\frac{16\nu^4 -40\nu^2 +9}{12a^2}
+\cos(2ka)\frac{(4\nu^2 -9)(-64 a^2 k^2+(1 -4\nu^2) ^2 )}{192 a^3
k} \bigg)+ \nonumber
\\
&&\mbox{Si}(2bk) \bigg( \cos(2ka)\frac{(4\nu^2 -9)(64 a^2 k^2-(1
-4\nu^2)^2 )}{192 a^3 k} -\sin(2ka)\frac{16\nu^4 -40\nu ^2
+9}{12a^2}  \bigg)\bigg\}
\end{eqnarray}
    \begin{figure}\label{de-nu}
    \epsffile{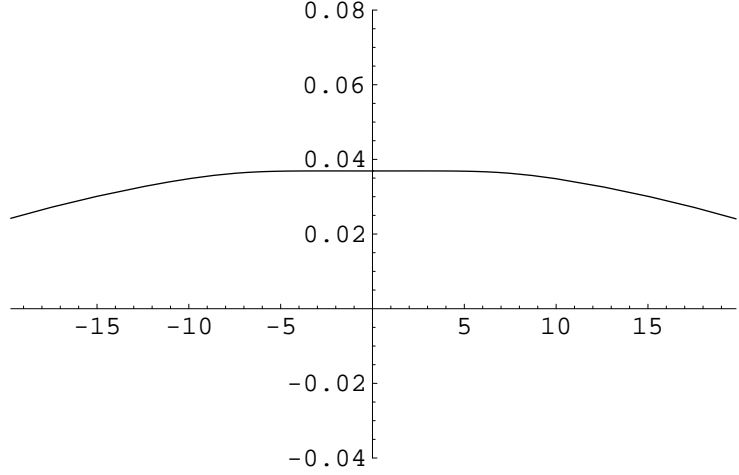}
    \caption{$\langle r^{-4} \rangle_\ell$ as a function of
    $\nu$, for $b/a = 10$ and $k\,a = 40$.}
    \end{figure}
where
\begin{eqnarray}
D(\nu,a,b,k)&=&8\,k\,\bigg( -4\,a\,( -1 + 4\,{\nu }^2 )  - \frac{(
a - b) \,( 64\,a^2\,k^2 + {( 1 - 4\,{\nu }^2) }^2 ) }{2} + 4\,a
\,( -1 + 4\,{\nu }^2) \,\cos (2\,( a - b) \,k) +  \nonumber
\\
&&\frac{( 64\,a^2\,k^2 - {( 1 - 4\,{\nu }^2) }^2 ) \,\sin (2\,( a
- b) \,k)}{4\,k}\bigg ) + \nonumber
\\
&&(4\nu^2-1)\bigg(\mbox{Ci}(2\,a\,k)\,( 16\,a\,k\,( -1 + 4\,{\nu
}^2 ) \,\cos (2\,a\,k) + ( 64\,a^2\,k^2 - {( 1 - 4\,{\nu }^2) }^2
) \,\sin (2\,a\,k) ))+\nonumber
\\
&&\mbox{Ci}(2\,b\,k)\,( -16\,a\,k\,( 4\,{\nu }^2 -1 ) \,\cos
(2\,a\,k) + ( -64\,a^2\,k^2 + {( 1 - 4\,{\nu }^2) }^2) \,\sin
(2\,a\,k)) - \nonumber
\\
&&[\mbox {Si}(2\,a\,k) - \mbox{Si}(2\,b\,k)]  [ ( 64\,a^2\,k^2 -
{( 1 - 4\,{\nu }^2 ) }^2 ) \,\cos (2\,a\,k) +16\,a\,k\,( 1 -
4\,{\nu }^ 2 ) \,\sin (2\,a\,k) ] \bigg)
\end{eqnarray}
and
\begin{eqnarray}
\mbox{Ci(z)}&=&- \int_z ^\infty \frac{\cos(t)}{t} dt \nonumber
\\
\mbox{Si(z)}&=& \int_0 ^z \frac{\sin(t)}{t} dt \nonumber
\end{eqnarray}

    Despite its aspect, $\langle r^{-4} \rangle$ is a slowly
    varying function of  $\nu$, as can be seen in Figure 1.
    Moreover, for a given $\nu$, $\langle r^{-4} \rangle$ rapidly
    approaches a constant value when $k$ grows up, as is shown in Figure
    2.

    Consequently, it is the coefficient of $\langle r^{-4}
    \rangle$ in (\ref{h1-val-med}),  the cubic polynomial
    $P(\ell,\Phi)$,
    which governs the shift produced on the
    eigenvalues. Notice that, for given flux $\Phi$
    and angular momentum $\ell$,
    the successive (large) eigenvalues of the radial equation
    (\ref{fer2}) are all shifted by the same constant. In
    particular, for large $|\ell|$, this constant does not change of sign.

    \begin{figure}\label{de-k}
    \epsffile{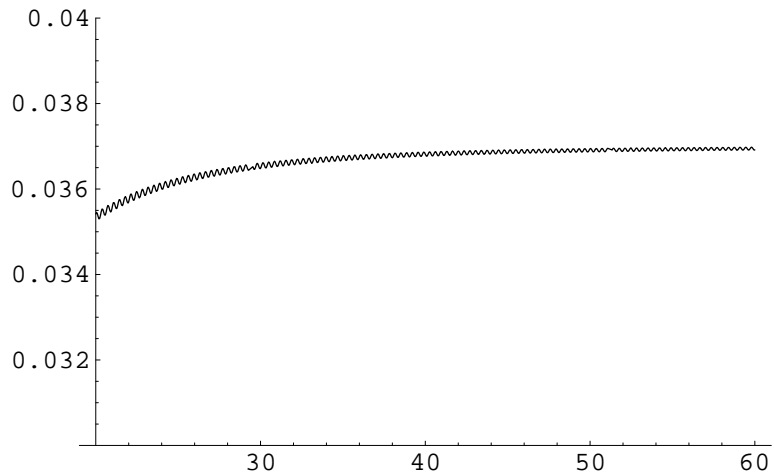}
    \caption{$\langle r^{-4} \rangle_\ell$ as a function of
    $k\, a$, for $b/a = 10$ and $\nu = 7-\sqrt{\pi}$.}
    \end{figure}

    Therefore, even though the 0-th order spectrum depends only on
    $\nu = |\ell-\Phi/2\pi|$, the first order
    ($\theta^1$) correction depends separately on the flux $\Phi$
    and the angular momentum $\ell$, introducing a shift in the
    eigenvalues sensitive to the sign of $\ell$. 
    
    Finally, we would to add some comments related to the relativistic case. Although in this paper we solve the Schr\"odinger equation, our
conclusions are valid in the relativistic case too, indeed, as the Aharonov-Bohm interaction is static, the Schr\"odinger and Klein-Gordon equations are related by $E_{Sch}\rightarrow E^2_{KG} -m^2$. However, a delicate point is the following; as we are thinking in electrons, one should use the Dirac equation instead of the Schrodinger one. In such case there is a critical subspace which admits non-trivial self-adjoint extensions 
\cite{pis,lajacky,ruso}.

In our case the boundary conditions ensures that the eigenfunctions have a finite limit for $r \rightarrow 0$. This could correspond to a possible self-adjoint extension. Any case, for first order corrections, as we have done, everything is consistent. For the perturbation (\ref{h11}), the problem is defined for $r \geq a$, which ruled out the case $r =0$. In spite of this constraint, one can consider the case $a \rightarrow 0$, but the boundary condition (\ref{fer33}) ensures the self-adjoint properties, as  {\it e.g. } in quantum mechanics.

\section{Scattering States for  the  Noncommutative Aharonov-Bohm Effect}

\subsection{The perturbative solution }
In order to compute the scattering states we look for solutions of
(\ref{schteta}) in the form
\begin{equation}
\Psi = \Psi_0 + \theta\, \Psi_1 + \dots , \nonumber
\end{equation}
    implying that
\begin{eqnarray}
\hat{H_0} \Psi _0 &=&  k^2 \Psi _0, \\
(\hat{H_0} - k^2)\Psi _1 &=& -\hat{H_1} \Psi _0. \label{ordenes}
\end{eqnarray}
    Therefore, the correction to the wave function at first order in
perturbation theory results
\begin{equation}
\Psi _1 (r,\varphi)= -(\hat{H_0} \,-\, k^2)^{-1} \hat{H_1}\Psi _0,
\label{hat}
\end{equation}
where the 0-th order wave function satisfies  the boundary conditions
\begin{eqnarray}
\Psi _0 (a,\varphi)&=&0, \nonumber
\\
\Psi _0 (r \rightarrow \infty,\varphi) &\sim& e^{i k r
\cos(\varphi)} + f(\varphi,k) \frac{e^{ikr}}{\sqrt{r}}.
\label{sammy}
\end{eqnarray}
The first equation guaranties that the electron never reaches the
region $r<a$, while the second one  is the usual scattering
condition.

The formal solution in (\ref{hat}) is given by
\begin{equation}
\Psi _1 (\mathbf{x})= - \int d\mathbf{x}'
G(\mathbf{x},\mathbf{x}') \hat{H_1}\Psi _0 (\mathbf{x}'),
\label{psi11}
\end{equation}
where $G(x,x')$ is the Green function of the unperturbed problem,
that is
\begin{equation}
(\hat{H_0} - k^2)G(r,\varphi;r',\varphi ')=\frac{1}{r}\,
\delta(r-r')\, \delta(\varphi-\varphi '). \label{green}
\end{equation}

\subsection{The Green Function}
We propose a solution for  (\ref{green}) of the form
\begin{equation}
G(r,\varphi;r',\varphi ')=\frac{1}{2 \pi} \sum _{\ell \in
Z}e^{i\ell (\varphi -\varphi ')} g_\ell (r,r'). \label{greeng}
\end{equation}
Replacing  this  in (\ref{green}) and using an appropriate
representation for the delta function, we obtain
\begin{equation}
(\hat{h}_{\ell,0} - k^2)g_\ell(r,r')=\frac{1}{r}\,\delta(r-r'),
\label{greengg}
\end{equation}
    where $g_\ell (r,r')$ must also satisfy the appropriate boundary
    conditions,
    \begin{equation}\label{bc-gl}
      g_\ell (a,r'>a) = 0, \quad g_\ell (r\rightarrow \infty,r')
      \sim \frac{e^{i k r}}{\sqrt{k r}}.
    \end{equation}

For $r\neq r'$, (\ref{greengg}) is just the Schr\"odinger equation
for the  commutative Aharonov-Bohm effect, whose solutions are
linear combinations of Bessel functions,
as in (\ref{fer3}). Let us introduce two linearly independent solutions of
    this homogeneous equation, satisfying the boundary condition at $r=a$
    and $r\rightarrow\infty$ respectively,
\begin{eqnarray}
\chi _\ell ^{(a)}(r)&=& Y_\nu (ka) J_\nu(kr) - Y_\nu(kr) J_\nu(ka), \nonumber \\
\chi _\ell ^{(\infty)}(r)&=&J_\nu(kr) + i
Y_\nu(kr)=H_\nu^{(1)}(kr), \label{jis}
\end{eqnarray}
    where $H_\nu^{(1)}(z)$ is the Hankel functions.

The  continuity of $g_\ell(r,r')$ at $r=r'$,
    together with the discontinuity in its first derivative
    implied by the RHS of (\ref{greengg}),
lead to
\begin{equation}\label{gder}
g_\ell(r,r')=C_0 \bigg \{\begin{array}{cc} \chi_\ell^{a}(r)\,
\chi_\ell ^{(\infty)}(r'), &\,\,\,\, r<r'
\\
\chi_\ell^{a}(r')\, \chi _\ell ^{(\infty)}(r) &\,\,\,\, r>r'
\end{array},
\end{equation}
    where the constant $C_0$ is given by
\begin{equation}
C_0=\frac{1}{r\, W[\chi_\ell ^{(a)}(r),\chi_\ell ^{(\infty)}(r)]},
\label{constant}
\end{equation}
being $W[f,g]=f\, g'-f'\, g $, the Wronskian.

\subsection{The free solution and the commutative case scattering theory}

The last ingredient we need for computing (\ref{psi11}) is to
express appropriately $\Psi_0$ satisfying the boundary conditions
(\ref{sammy}). We write
\begin{equation}
\Psi _0(r,\varphi)=\sum _{\ell \in Z} e^{i\ell \varphi}\,
\chi_\ell^{(0)}(r), \label{cero}
\end{equation}
with
\begin{equation}\label{chil0}
  \chi_\ell^{(0)}(r) =\bigg[
    A_{\ell} J_\nu(k r) + B_\ell Y_\nu (k r)\bigg].
\end{equation}

For convenience, in what follows we will develop a partial waves
analysis of the scattering amplitude, as in \cite{rui}. There are
other treatments of this problem in the literature (see, for
example, \cite{ab}, \cite{hagen} and \cite{soldati}) leading to
results differing in the forward scattering term, but having the
same scattering amplitude for non vanishing angles. This justifies
our approach to the cross section for $\varphi \neq 0$.

The first condition in (\ref{sammy}) implies that
\begin{equation}
A_{\ell} J_\nu(k a) + B_\ell Y_\nu (k a)=0. \label{albl}
\end{equation}
In the second condition, one can develop in Fourier series the
scattering amplitude $f(\varphi,k) = \sum_{\ell\in Z} e^{i \ell
\varphi} f_\ell$, and the plane wave $e^{i k x}$ (which  can be
written in terms of Bessel functions):
\begin{eqnarray}
e^{i k r \cos(\varphi)} + f(\varphi,k)\frac{ e^{ikr}}{\sqrt{r}}
&=&\sum _{\ell \in Z} e^{i\ell \varphi} \bigg[i^{|\ell|}
J_{|\ell|}(kr) + f_\ell \frac{ e^{ikr}}{\sqrt{r}}\bigg] \nonumber
\\
&\sim&\sum _{\ell \in Z} e^{i\ell
\varphi}\bigg[\frac{e^{ikr}}{\sqrt{r}}\left(\frac{i^{|\ell|}}
{\sqrt{{2 \pi k }}} \, e^{-i(\frac{\pi|\ell|}{2}+\frac{\pi}{4})}
+f_\ell\right) + \frac{e^{-ikr}}{\sqrt{r}}\left(
\frac{i^{|\ell|}}{{\sqrt{{2 \pi k }}}} \,
e^{i(\frac{\pi|\ell|}{2}+\frac{\pi}{4})}\right)\bigg],
\label{expan}
\end{eqnarray}
where we have replaced the asymptotic expression of Bessel
functions inside the series.

Comparing the terms in (\ref{expan}) with the asymptotic
expression of $\chi_\ell^{(0)}(r)$ in (\ref{cero}) for large
values of $k\, r$ (see equation (\ref{fer4})) we get the following
equations
\begin{eqnarray}
i^{|\ell|}e^{-i(\frac{\pi|\ell|}{2}+\frac{\pi}{4})} +\sqrt{2 \pi
k}f_\ell&=& \left( A_{\ell} -i B_{\ell} \right)
e^{-i(\frac{\pi\nu}{2}+\frac{\pi}{4})}, \label{set1}
\\
i^{|\ell|}e^{i(\frac{\pi|\ell|}{2}+\frac{\pi}{4})}&=&\left(
A_{\ell} + i B_{\ell}\right) e^{i(\frac{\pi\nu}{2}+\frac{\pi}{
4})}. \label{set2}
\end{eqnarray}

The solution to the set of equations  (\ref{albl}),  (\ref{set1})
and (\ref{set2}) is
\begin{eqnarray}
A_\ell&=&i\, e^{- \frac{i }{2}\,\pi \,\left( \nu  - 2\,\ell
\right)}\, \frac{ Y_\nu(ka )}{ H^{(1)}_\nu (ka)}, \label{palma1}
\\
B_\ell&=&-i\, e^{-\frac{i }{2}\,\pi \,\left( \nu  - 2\, \ell
\right)}\, \frac{J_\nu (ka)}{H^{(1)}_\nu (ka)}, \label{palma2}
\\
f_\ell&=&- \frac{e^{-i\frac{\pi}{4}}}{\sqrt{2\pi k}}\left[1
+e^{-i\pi(\nu-\ell)} \frac{H^{(2)}_\nu (ka)}{H^{(1)}_\nu (ka)}
\right], \label{palma3}
\end{eqnarray}
where the $H^{(1,2)}_\nu(z)$ are the Hankel functions.

>From (\ref{palma3}) one can easily extract  the phase shifts.
Indeed, from scattering theory \cite{taylor}, one knows that the
scattering amplitude for the $\ell$-th partial wave is
\begin{equation}
f_\ell = \frac{e^{-i\frac{\pi}{4}}}{\sqrt{2\pi k}} \left( e^{2i
\delta_\ell} -1\right). \nonumber
\end{equation}
Then, in the present case
\begin{equation}
e^{2i \delta_\ell} =  {(-1)}^{\ell+1} e^{-i\pi \nu}
\frac{H^{(2)}_\nu (ka)}{H^{(1)}_\nu (ka)}, \label{fase1}
\end{equation}
   which provides an exact expression for the $S$-matrix \cite{rui}.

One can check the consistency of our approach by evaluating the
limit $a \rightarrow 0$.  In this case $f_\ell$ reduces to
\begin{equation}
f_{0,\ell} = \frac{e^{-i\frac{\pi}{4}}}{\sqrt{2\pi
k}}\left[e^{-i\pi(\nu-\ell)} -1 \right], \label{fase2}
\end{equation}
or, equivalently,
\begin{equation}
\delta_\ell =\frac{\pi}{2} \left( |\ell| -\nu\right).
\label{fase3}
\end{equation}
Equations (\ref{fase1})-(\ref{fase3}) are in agreement with other
derivations found in the literature \cite{rui,jackiw}. Notice that
the phase shifts $\delta_\ell$ do not tend to 0 for
$\ell\rightarrow \pm \infty$; instead, they approach to non
vanishing constants (see the discussion in \cite{hagen}).

In order to compute the differential cross section one must get,
firstly, the total scattering amplitude, {\it i.e.} we must
evaluate the sum
\begin{equation}
f(\varphi,k) = \sum_{\ell=-\infty}^\infty f_\ell \, e^{i\, \ell
\,\varphi}. \label{amp1}
\end{equation}

The explicit calculation of (\ref{amp1}) involves several
technical and conceptual difficulties which have been source of
controversy in the past \cite{conf}.

Firstly, let us consider the case $a=0$. Making use of
(\ref{fase2}), the total amplitude becomes
\begin{equation}
f_0(\varphi,k) =  \frac{e^{-i\frac{\pi}{4}}}{\sqrt{2\pi k}}\,
\sum_{\ell=-\infty}^\infty\, e^{i\ell \varphi} \left( -1+ e^ {-i
\pi \nu} \,{(-1)}^{\ell}\right). \label{amp2}
\end{equation}

The first term in (\ref{amp2}) is
\begin{equation}
\frac{e^{-i\frac{\pi}{4}}}{\sqrt{2\pi k}}\,
\sum_{\ell=-\infty}^\infty\, e^{i\ell \varphi} \left( -1\right) =
-\frac{e^{-i\frac{\pi}{4}}}{\sqrt{2\pi k}}\, 2 \pi\, \delta [
\varphi]  = - \sqrt{\frac{2 \pi}{i k}} \, \delta [ \varphi].
\label{amp3}
\end{equation}

For the second one we get (see Appendix B for details)
\begin{equation}
\frac{e^{-i\frac{\pi}{4}}}{\sqrt{2\pi k}}\,
\sum_{\ell=-\infty}^\infty\, e^{i\ell \varphi} \left( e^ {-i \pi
\nu} \,{(-1)}^{\ell}\right)= \frac{e^{-i\frac{\pi}{4}}}{\sqrt{2\pi
k}}\, \left\{2\pi\cos\left(\frac{\Phi} 2\right)\delta[\varphi] +
2i\sin\left(\frac{\Phi}2\right)\,{\cal
P}\bigg[\displaystyle{\frac{e^{i(\ell_0+1)
\varphi}}{1-e^{i\varphi}}}\bigg]\right\}, \label{amp4}
\end{equation}
where $l_0$ is the integer part of $\frac{\Phi}{2\pi}$ and ${\cal
P}[F(\varphi)]$ denotes the principal value of $F(\varphi)$.

Finally, the scattering amplitude becomes
\begin{eqnarray}
f_0(\varphi,k) &=& \sqrt{\frac{2 \pi}{i k}} \left\{ \left[
\cos\left(\frac{\Phi}{2}\right) -1\right] \,\delta [\varphi] +
\frac{i}{\pi} \sin\left( \frac{\Phi}{2}\right)\,{\cal P}\left[
\frac{e^{i(l_0 +1)\varphi}}{1 - e^{i\varphi}}\right]
\right\}\nonumber
\\
&=& \sqrt{\frac{2 \pi}{i k}} \left\{ \left[
\cos\left(\frac{\Phi}{2}\right) -1\right] \,\delta [\varphi] +
\frac{i}{\pi} \sin\left( \frac{\Phi}{2}\right)\,\left( {\cal
P}\left[ \frac{i}{\varphi}\right] + \left[\frac{e^{i(l_0
+1)\varphi}}{1 - e^{i\varphi}} -
\frac{i}{\varphi}\right]\right)\right\}. \label{amp5}
\end{eqnarray}
Notice that $f_0(\varphi,k)$ vanishes for $\Phi = 4 \pi n$, with
$n$ integer. For these flux values the particles are not scattered
at all by the zero radius solenoid in the commutative case.

This formula coincides with equation (4.11) of \cite{rui}, where
it  was obtained following a different procedure. The
interpretation of the forward scattering term in (\ref{amp5}), in
the context of the construction of the scattering matrix, is
considered in that reference. Many authors have discussed the
presence or not of this forward scattering singular term in the
total scattering amplitude (see, for example \cite{hagen}). It is
not present in the original derivation by Aharonov and Bohm
\cite{ab}, and can be also avoided making use of an analytic
regularization as in \cite{soldati}. However, as previously
pointed out, in the present work we are interested in the
calculation of the differential cross section for scattering
angles different from zero, where different approaches coincide.
This justifies the partial waves analysis we performed.

The calculation of the differential cross section is now
immediate. Indeed, for $\varphi \neq 0$ we have
\begin{equation}
\frac{d\sigma}{d\varphi} = |f_0(\varphi,k)|^2 = \frac{\sin^2
[\frac{\Phi}{2} ]}{2 \pi k \,\sin^2 [\frac{\varphi}{2}]},
\label{cs}
\end{equation}
which is the usual Aharonov-Bohm differential cross section
\cite{ab}, vanishing for $\Phi = 2 \pi n$, with $n$ integer.

If the radius of the solenoid is different from zero ($a>0$), one
    can similarly isolate the singular contributions to the total scattering
    amplitude $f(\varphi,k)$, coming
    from large values of $\ell$ (or equivalently, from large values of
    $\nu$). Using appropriate large order expansions for the Hankel
    functions, one finds that the coefficient
    $f_\ell$ is given in this case by the RHS of (\ref{fase2})
    plus terms rapidly decreasing with $\ell$, which lead to
    absolutely convergent series (summing up to continuous
    functions of $\varphi$). Therefore, the singular terms found in
    (\ref{amp5}) for $f_0(\varphi,k)$ (those containing $\delta [\varphi]$
    and ${\cal P}\left[ \frac{i}{\varphi}\right]$) are also present in
    $f(\varphi,k)$.

\section{First order noncommutative corrections to the scattering amplitude}

In this section we calculate  the first order ($\theta^1$)
perturbative correction to the scattering amplitude
$f(\varphi,k)$. This will allow to find the first noncommutative (singular) corrections to the
differential cross section.

In doing so, we must evaluate $\Psi_1(\mathbf{x})$ in
(\ref{psi11}), with $\Psi_0(\mathbf{x})$ given in (\ref{cero}),
(\ref{palma1}) and (\ref{palma2}), and $G(\mathbf{x},\mathbf{x}')$
given in (\ref{greeng}), (\ref{gder}) and  (\ref{constant}).

Taking into account that $\hat{H}_0$ and $\hat{H}_1$ are diagonal
in $\ell$, we can write $\Psi_1(r,\varphi) = \sum_\ell e^{i \ell
\varphi}\, \chi_\ell^{(1)}(r)$ to get
\begin{equation}
    \begin{array}{c}
      \chi_\ell^{(1)}(r) = - \int_a^\infty g_\ell(r,s)\,
      \left(\hat{h}_{\ell,1}\chi_\ell^{(0)}(s)\right) \, s\, ds =
     \\ \\
       = - C_0 \left[ \chi^{(\infty)}_\ell
    (r) \int_a^r  \chi^{(a)}_\ell (s)\,\left( {\hat h_{\ell,1}}
    \chi^{(0)}_\ell (s)\right)\,s\,ds + \chi^{(a)}_\ell (r) \int_r^\infty
    \chi^{(\infty)}_\ell (s) \,\left( {\hat h_{\ell,1}}
    \chi^{(0)}_\ell (s)\right)\,s\,ds \right],
    \label{or3}
    \end{array}
\end{equation}
with $\chi^{(\infty)}_\ell(r)$ and $\chi^{(a)}_\ell(r)$ given in
(\ref{jis}), and $ \hat{h}_{\ell,1}$ given in (\ref{h11}).

Since we are interested in the noncommutative corrections to the
scattering amplitude, we should consider the asymptotic behavior
of $\chi_\ell^{(1)}(r)$ for $r\rightarrow \infty$. The expansions
for large arguments of Bessel function in (\ref{fer4}) allows to
see that the second term into the brackets in the RHS of
(\ref{or3}) decreases faster than the first one, and can be
discarded.

For arbitrary $a>0$, the integrand in the first term  is too
complicated to give a closed solution to this integral, and some
simplification is necessary. For this reason, we will analyze it only in
the $a\rightarrow 0$ limit.

    In this limit, straightforward calculations lead to
\begin{equation}
C_0 \chi^{(a)}_\ell (r) \rightarrow_{a\rightarrow 0} -\frac{i
\pi}{2} J_\nu (k r), \label{si1}
\end{equation}
and
\begin{eqnarray}
{\hat h_{\ell,1}}  \chi^{(0)}_\ell (s)& \rightarrow_{a\rightarrow
0}&  \frac{{(-1)}^\ell \, e^{-\frac{1}{2} i \pi \nu} }{8 \pi^3
s^4} \biggl\{ \pi^2\, ( 4 \pi \ell  - \Phi)\,k\,s\,\left(
J_{\nu+1} (k s)-J_{\nu-1} (k s) \right)+\nonumber
\\
&&+ \left(  8 \pi^3 l^3 - 8 \pi^2 \ell^2 \Phi + 4 \pi \ell \Phi^2
- \Phi^3\right) J_\nu (k s) \biggr\}, \label{si2}
\end{eqnarray}
while $\chi_\ell^{(\infty)}(r)$ does not depend on $a$.

Then, for $\nu>1$ \footnote{For $\nu\leq 1$, the first integral in
the RHS of (\ref{or3}) is divergent in the $a\rightarrow 0$ limit,
and one should keep $a>0$ in these two (or at most three) terms.
For the consideration of the most general acceptable boundary
conditions at $r=0$, see for example the study of the self adjoint
extensions of the Hamiltonian  discussed in \cite{soldati}. Since
we are actually interested in the singular contributions to the
scattering amplitude for small angles, coming from the large
$|\ell|$ behavior of $\chi_\ell^{(1)}(r)$, we will not worry about
these terms (whose contributions to $\Psi_1$ are smooth,
continuous function of $\varphi$).} the coefficient of
$\chi_\ell^{(\infty)}(r)$ in the RHS of (\ref{or3}) reduces in
this approximation to
\begin{equation}
 \frac{i}{16 \pi^2}  \, {(-1)}^\ell e^{-\frac{i}{2} \pi \nu} \,
k^2 \, \left[ \frac{( 8 \pi^3\ell^3 - 8 \pi^2  \ell^2 \Phi- 8
\pi^3 \ell
 + 4 \pi \ell  \Phi^2 + 2 \pi^2 \Phi- \Phi^3 )}{4 \nu
(\nu^2-1)} \right], \label{si3}
\end{equation}
expression in which we must distinguish two cases, namely
 \begin{equation}
 \ell>\frac{\Phi}{2 \pi} \quad {\mbox{and}} \quad
 \ell \leq \frac{\Phi}{2 \pi}.  \nonumber
 \end{equation}

If $\ell \geq  l_0  + 1 >  \frac{\Phi}{2 \pi}$ (where $l_0$ is the
integer part of  $\frac{\Phi}{2 \pi}$), then
$\nu=\ell-\frac{\Phi}{2 \pi}$, and (\ref{si3}) becomes
 \begin{equation}
 i \,{(-1)}^\ell \,e^{-\frac{i}{4} ( 2 \pi \ell  - \Phi)} \pi\, k^2
 \left( \frac{1}{8} + \frac{\Phi}{16 \pi \ell} + {\cal O}(\ell^{-2}) \right). \label{in1}
 \end{equation}

Now, taking into account that, for $r \rightarrow \infty$,
 \begin{equation}
 \chi^{(\infty)}_\ell(r)=
 H_\nu^{(1)}(kr) \sim \sqrt{\frac{2}{\pi k}}\, e^{-i\frac{\pi }{2}(\ell-\frac{\Phi}{2\pi}) -
 i \frac{\pi}{4}} \,\frac{e^{i k\,r}}{\sqrt{r}}, \label{w}
 \end{equation}
we get the first perturbative correction to $f_\ell$ in
(\ref{amp1}) as
 \begin{equation}
 \theta\, f_{1,\ell}= \frac{\theta}{4}\, \sqrt{\frac{i \pi}{2}} \,e^{
 i \frac{\Phi}{2}} \,\,  k^{3/2} \,\, \left( 1 + \frac{\Phi}{2 \pi \ell} +
 {\cal O}(\ell^{-2})\right).
 \label{ww}
 \end{equation}

Multiplying this expression by $e^{i \ell \varphi}$ and summing on
$\ell$ from $\ell_0 +1$ to $\infty$, we obtain the following
contribution to the scattering amplitude:
 \begin{equation}
 \frac{\theta}{4}\sqrt{\frac{i\pi}{2}} \,e^{
 i \frac{\Phi}{2}}\,\,  k^{3/2} \,\,\left( \pi\, \delta [\varphi] + {\cal P}\left[
 \frac{1}{1-e^{i \varphi}} \right]
 - \frac{\Phi}{2 \pi} \log [1 -  e^{i(i \epsilon + \varphi)}]+\dots \right), \label{a1}
 \end{equation}
where the $\epsilon \rightarrow 0^+$ limit is understood, and the
dots stand for continuos functions of $\varphi$.

For the case $\ell \leq \ell_0 \leq \frac{\Phi}{2 \pi}$, we have $
\nu =  \frac{\Phi}{2 \pi}-\ell$, and a similar calculation (where
the sum on $\ell$ is taken from $-\infty$ to $\ell_0$) leads to
the following contribution to the scattering amplitude:
\begin{equation}
\frac{\theta}{4}\sqrt{\frac{i\pi}{2}} \,e^{ - i
\frac{\Phi}{2}}\,\,  k^{3/2} \,\,\left( \pi\, \delta [\varphi] -
{\cal P}\left[ \frac{1}{1-e^{-i \varphi}} \right]
 - \frac{\Phi}{2 \pi} \log [1 -  e^{-i(-i \epsilon + \varphi)}]
 +\dots\right). \label{a2}
 \end{equation}
where again, the limit $\epsilon \rightarrow 0^+$ is understood
and the dots represent continuos functions of $\varphi$.

Therefore, at first order in $\theta$ the scattering amplitude is
corrected by the addition of
 \begin{eqnarray}
\theta\,  f^{(1)} (\varphi,k) &=& \frac{\theta}{4}
\,\sqrt{\frac{i\pi}{2}} \, k^{3/2}
 \, \biggl\{ 2 \pi \cos\left(\frac{\Phi}{2}
 \right) \,\delta \left [\varphi\right]+
 i\, \cos\left( \frac{\Phi - \varphi}{2}\right) \, {\cal P}
 \left[ \frac{1}{\sin \left( \frac{\varphi}{2}\right)}\right]
 - \nonumber \\
 &-&\frac{ \Phi}{2 \pi}\,e^{- i \frac{\Phi}{2}} \,  \log [ 1- e^{-i(-i
\epsilon +  \varphi)}]  -\frac{ \Phi}{2 \pi}\, e^{ i
\frac{\Phi}{2}} \, \log [ 1- e^{i(i \epsilon + \varphi)}] + \dots
\biggr\}. \label{a3}
 \end{eqnarray}

In conclusion, as the incident particles are very energetic and
the scattering angle is very small, the main contributions to the
total scattering amplitude $f(\varphi,k)$ are given by
\footnote{The reader could note here that this argument is
reminiscent of the eikonal approximation, see {\it e.g.} J. J. Sakurai, {\it Modern Quantum Mechanics}, Addison Wesley (1994).}
 \begin{eqnarray}
 f(\varphi,k) &=& \left\{ \sqrt{\frac{2\pi}{i k}}
 \left[ \cos \left(\frac{\Phi}{2}\right)-1 \right] +\theta\,\frac{\pi}{2}\sqrt{\frac{i\pi}{2}}
 { \, k^{3/2} \,\cos \left(\frac{\Phi}{2}\right)} \right\} \delta
 [\varphi]+
 \\
 &+& \left\{ -\sqrt{\frac{2}{i \pi k}}
 {\, \sin \left(\frac{\Phi}{2}\right)} + \theta\, \frac{i}{2}
 \,\sqrt{\frac{i \pi}{2}}  \,  k^{3/2}
 \, \cos\left(\frac{\Phi}{2}\right) \right\}
 {\cal P} \left[\frac{1}{\varphi}\right] + \nonumber
 \\
 &{-}& \frac{\theta}{4}\,\sqrt{\frac{i}{2\pi}} \, {\,k^{\frac{3}{2}}\,\Phi \,}
 \cos \left(\frac{\Phi}{2}\right) \, \log (\varphi)+
 {\mbox{continuous}} \,\,{\mbox{functions}}\,\,{\mbox{of}}\,\,
 \varphi.
 \label{a4}
 \end{eqnarray}
Notice that the most singular terms in the scattering amplitude,
which are $\sim k^{-1/2}$, are corrected by noncommutative terms
$\sim \theta\, k^{3/2}$. Moreover, for $\Phi=4\pi n$, with $n$
integer, the 0-th order singular terms in the amplitude vanish,
contrarily to the noncomutative corrections, which are different
from zero.

For small angles $\varphi\neq 0$, the dominant term in the
amplitude is $\sim 1/\varphi$. Then, for the differential cross
section we have
\begin{equation}\label{dif-cross}
  \frac{d \sigma}{d \varphi} = \left\{ \frac{2}{\pi k}\,
  \sin^2\left(\frac{\Phi}{2}\right)+
  \theta\, \frac{k}{2}\, \sin\left(\Phi\right)+
  \theta^2\, \frac{\pi}{8}\,k^3\,\cos^2\left( \frac{\Phi}{2} \right)
  \right\}\frac{1}{\varphi^2} + {\rm less}\ {\rm singular}\ {\rm
  terms}.
\end{equation}

Now, if the magnetic flux is quantized as $\Phi=2 \pi n$, with $n$
integer, the differential cross section at small angles is
dominated by noncommutative effects,
  \begin{equation}
  \frac{d\sigma}{d\varphi} = \theta^2 \, \frac{\pi k^3 }{8\varphi^2}
  + {\rm less}\ {\rm singular}\ {\rm
  terms}. \label{se1}
  \end{equation}
It is interesting to note that, contrarily to the usual
Aharonov-Bohm effect,  in the noncommutative case the differential
scattering cross section is different from zero when the magnetic
flux is quantized.

 Apparently, this correction ($\sim \theta^2$)
could be relevant at high energies. This simple formula will allow
us to extract interesting physical information, as we will see in
the next section.

\section{Phenomenological estimations for spatial noncommutative effects}

As mentioned in the introduction, the Aharonov-Bohm effect is an
important  mechanism to explain other physical phenomena. This
point of view has been used in the past, and some applications of
this idea are cosmic strings and GUT \cite{aw}, anyons\cite{fw2}
and also three-dimensional gravity\cite{jackiw}.

In this section we will analyse experimental possibilities of
detecting noncommutative signals via the Aharonov- Bohm effect.
Our numerical estimations --as we will see below-- show that these
relics could be explored in particle physics experiments involving
energies between 200 and 300 GeV, if the present bound for
$\theta$ is correct.

In order to estimate a bound for the $\theta$ parameter, first we
note that, since noncommutative effects are tiny, the corrections
to the differential cross section could be, typically, of the
order of the cross section for neutrino events $\sim
10^{-3}\,\,$nb. If we  choose the scattering angle between 1 and 2
degrees, and  take an energy $\sim$200 GeV as the highest possible
presently available for electrons, then we find:

\begin{equation}\label{kkk}
\theta =  \left[ \left(\frac{8 \varphi^2}{\pi k^3}\right)
\frac{d\sigma}{d\varphi}\right]^{\frac{1}{2}} \sim  [
10 \,\mbox{TeV}]^{-2},
\end{equation}
which is in agreement with the bound given in \cite{carroll}.

Thus,  precise measurements of the differential cross section for
small angles could give us information about spatial
noncommutativity.

\section{Conclusions}

Three relevant properties of the remarkable phenomenon of
noncommutative Aharonov-Bohm effect have been found in the present
paper:

$\bullet$ Pattern fringes can appear even when the magnetic flux
is quantized, contrarily to the commutative case.

$\bullet$ The differential cross section, given by (\ref{se1}), is
different from zero when the magnetic flux is quantized.

$\bullet$ Our results allow for an estimation of a bound for the
noncommutative parameter $\theta$, which is in agreement with
\cite{carroll}.

The first property, in principle, could be verified in a Tonomura
like experiment, if an appropriate incident electron beam is
available. Our estimations suggest, however, that the incident
electron beam energy should be much larger than the energy reached
in these experiments \cite{tono}. Thus, an experimental
verification should be searched  in high energy physics
experiments and, specially, by measuring differential cross
sections for small angles.

\acknowledgments This work has been partially supported by the
grants 1010596, 1010976, 7010976 and 3000005 from Fondecyt- Chile,
and also by the grant  C-13398/6 from Fundacion Andes.

HF also acknowledge support from CO\-NI\-CET (grant 0459/98) and
UNLP (grant 11/X298), Argentina.

\vskip 1.0cm
\appendix
\section{Note on the Relativistic Aharonov-Bohm Effect}
In this appendix, we would like to discuss some implications of
the relativistic Aharonov-Bohm effect.

    From reference \cite{vitorio} one can see that the
Green function associated to the usual Aharonov-Bohm effect is
given by
\begin{equation}
G[x,x^{'}] = \sum_{n=-\infty}^{\infty} (-i)^{|n+\varphi|}
\exp[-i(n+\Phi)] F_{|n+\Phi|}, \label{gre1}
\end{equation}
where $\Phi$ is the magnetic flux and the function $F_{|n+\Phi|}$
for the non-relativistic case is
\begin{equation}
F_{|n+\Phi |}= \frac{m}{2\pi i} \exp[ \frac{2m i}{\tau} (r^2 +
r^{'2})] J_{|n+\Phi|} ( \frac{mrr^{'}}{\tau}), \label{nr}
\end{equation}
where $\tau = t-t^{'}$ and $J_{\alpha }$ are Bessel functions. For
the the relativistic case the calculation is similar. Indeed,
after using the proper-time gauge the function $F_{|n+\Phi |}$
becomes
\begin{eqnarray}
F_{|n+\Phi |}&=& \nonumber
\\
\int d^2p \int_0^\infty dT \exp[&i& p_\mu \Delta x^\mu -
\frac{T}{2} (p^2 +m^2)] J_{|n+\Phi|} ( \frac{r r^{'}}{T}).
\label{r}
\end{eqnarray}
where $T= N(0) (t-t^{'})$ with $N(0)$ the einbein.

If we use the Poisson summation formula, then in both the
relativistic as well as in the non relativistic case, the Green
function is
\begin{equation}
G[x,x^{'}] = \sum_{n=-\infty}^{\infty}  e^{2i\pi n \Phi} K_n,
\label{ho}
\end{equation}
where $K_n$ is defined as
\begin{equation}
K_n = \int_{-\infty}^\infty d\omega\,(-i)^{|\omega|}\, e^{-i\omega
\Phi}\, F_{|\omega|}, \nonumber
\end{equation}
and, as a consequence, the wave function becomes
\begin{equation}
\psi (x) = \sum_{n=-\infty}^{\infty} e^{2i\pi n \Phi} \varphi_n
(x), \label{wf}
\end{equation}
with
\begin{equation}
\varphi_n (x) = \int \,dy\, \,G_n\,[\,x,y]\, \psi (y),
\end{equation}
being $\varphi_n$ and $G_n [\,x,y]$, respectively, the wave and
Green functions for the n-th homotopy class \cite{dewitt}.

Thus, from (\ref{wf}) one see that the relativistic character of
the system is contained in $K_n$ and only  the exponential factor,
which does not depend on the energy, is responsible for the fringe
pattern. This result reflects the topological nature of the
commutative Aharonov-Bohm effect. However, our formula
(\ref{segundo}) shows us that the noncommutative Ahararonov-Bohm
effect is radically different  because the fringe pattern must
change when the electrons are getting higher energies.

\section{Derivation of equation (\ref{amp4})}

In this appendix  we show that
\begin{equation}
\sum_{\ell=-\infty}^{\infty}  e^{i\ell\varphi}e^{-i\pi \nu}
{(-1)}^\ell =2\pi\cos\left(\frac{\Phi} 2\right)\delta[\varphi] +
2i\sin\left(\frac{\Phi}2\right)\,{\cal
P}\bigg[\displaystyle{\frac{e^{i(\ell_0+1)
\varphi}}{1-e^{i\varphi}}}\bigg], \label{iden1}
\end{equation}
where $\ell_0$ is the integer part of $\Phi/2\pi$.
    Firstly, notice that $\displaystyle{e^{-i\pi \nu} {(-1)}^\ell = e^{i \pi ( |\ell| -|\ell -\frac{\Phi}{2\pi}|)}},$
    since the exponents coincide modulo $2 \pi$.  Moreover,
if $\ell\geq \ell_0+1 $ then $|\ell-\frac{\Phi}{2\pi}| =
\ell-\frac{\Phi}{2\pi}$, while if  $\ell \leq \ell_0$ then
$|\ell-\frac{\Phi}{2\pi}|=-\ell + \frac{\Phi}{2\pi}$.

Therefore, we can split the series in (\ref{iden1}) to write
\begin{eqnarray}
\sum_{\ell=-\infty}^{\infty}  e^{i\ell\varphi}e^{i \pi ( |\ell| -
|\ell - \frac{\Phi}{2\pi}|)}&=&\sum_{\ell=-\infty}^{\ell_0}
e^{i\ell \varphi}e^{- i \pi (
\frac{\Phi}{2\pi})}+\sum_{\ell=\ell_0+1}^{\infty}
e^{i\ell\varphi}e^{i \pi ( \frac{\Phi}{2\pi})} \nonumber
\\
&\equiv&e^{-i\frac{\Phi}{2}}e^{i\ell_0\varphi}\lim_{\epsilon\rightarrow
0} \sum_{\ell=0}^{\infty} e^{-i\ell(\varphi-i\epsilon)}
+e^{i\frac{\Phi}{2}}e^{i(\ell_0+1)\varphi}\lim_{\epsilon\rightarrow
0} \sum_{\ell=0}^{\infty}  e^{i \ell(\varphi+i\epsilon)},
\label{sum2}
\end{eqnarray}
where we have introduced the positive parameter $\epsilon$ to
properly define these sums.

Now, the evaluation goes in the standard way. For the first series
we have
\begin{eqnarray}
e^{-i\frac{\Phi}{2}}e^{i\ell_0\varphi}\lim_{\epsilon\rightarrow 0}
\sum_{\ell=0}^{\infty}  e^{-i\ell(\varphi-i\epsilon)}&=
&\lim_{\epsilon\rightarrow 0}
\frac{e^{-i\frac{\Phi}{2}}e^{i\ell_0\varphi}}
{1-e^{-i(\varphi-i\epsilon)}} \nonumber
\\
&=&-i e^{-i\frac{\Phi}{2}}e^{i(\ell_0+1)
\varphi}\bigg(i\pi\,\delta[i(1-e^{i\varphi})] + {\cal
P}\bigg[\frac{1}{i(1-e^{i\varphi})}\bigg]\bigg) \nonumber
\\
&=&\pi e^{-i\frac{\Phi}{2}}\, \delta[\varphi] -
e^{-i\frac{\Phi}{2}} \, {\cal
P}\bigg[\frac{e^{i(\ell_0+1)\varphi}}{1-e^{i\varphi}} \bigg],
\end{eqnarray}
    where ${\cal P}[\dots]$ means principal value.

The second series in the RHS of (\ref{sum2}) is evaluated in a
similar way
\begin{eqnarray}
e^{i\frac{\Phi}{2}}e^{i(\ell_0+1)\varphi}\lim_{\epsilon\rightarrow
0} \sum_{\ell=0}^{\infty}  e^{i\ell(\varphi+i\epsilon)}&=&
\lim_{\epsilon\rightarrow 0}
\frac{e^{i\frac{\Phi}{2}}e^{i(\ell_0+1)\varphi}} {1-
e^{i(\varphi+i\epsilon)}} \nonumber
\\
&=& i e^{i\frac{\Phi}{2}}e^{i(\ell_0+1) \varphi} \bigg(-
i\pi\,\delta[i(1-e^{i\varphi})] + {\cal
P}\bigg[\frac{1}{i(1-e^{i\varphi})} \bigg]\bigg) \nonumber
\\
&=&\pi e^{i\frac{\Phi}{2}} \,\delta[\varphi] +
e^{i\frac{\Phi}{2}}\, {\cal
P}\bigg[\frac{e^{i(\ell_0+1)\varphi}}{1-e^{i\varphi}} \bigg].
\end{eqnarray}

Collecting both results one finally obtains
\begin{equation}
\sum_{\ell=-\infty}^{\infty}  e^{i\ell\varphi}e^{-i\pi \nu}
{(-1)}^\ell = 2\pi\cos\left(\frac{\Phi} 2\right)\delta[\varphi] +
2i\sin\left(\frac{\Phi}2\right)\,{\cal
P}\bigg[\displaystyle{\frac{e^{i(\ell_0+1)
\varphi}}{1-e^{i\varphi}}}\bigg]
 . \label{idenfinal}
\end{equation}


\begin{references}
\bibitem{string} A. Connes, M. Douglas and A. S. Schwarz, {\it JHEP}
9802:003 (1998);
N. Seiberg and E. Witten, {\it JHEP} {\bf 09}, 032 (1999).
\bibitem{various} The literature is very extense, some references are;
T. Filk, {\it Phys.Lett.} {\bf B376}, 53 (1996); R.
Gopakumar, J. Maldacena, S. Minwalla and A. Strominger, {\it JHEP}
{\bf 0006} 036 (2000); L. Alvarez-Gaum\'e and S. Wadia,
hep-th/0006219; M. Hayakawa, {\it Phys. Lett.} {\bf 476}, 431
(2000); C. P. Martin and F. Ruiz, hep-th/0007131;  A.  Armoni, {\it Nucl.Phys.} {\bf  B593} (2001); J. Gomis and T.
Mehen, hep-th/0005129; I. Mociou, M. Popelov and R. Roiban, {\it
Phys. Lett.} {\bf 489B}, 390 (2000);  C.E. Carlson, C.D. Carone and R.F. Lebed, {\it Phys. Lett.} {\bf B518}, 201 (2001); K. Valavano, hep-th/0006245;
C. Duval and P. A. Horv\'athy, hep-th/0106089 and {\it ibid}, {\it
Phys. Lett.}{\bf 479B}, 284 (2000);  Z. Guralnik, R. Jackiw, S.Y.
Pi and A.P. Polychronakos, hep-th/0106044;  M. Lubo,
hep-th/0106018.; I. Mocioiu, M. Popelov and R. Roiban,
hep-ph/0005191;  J. Gamboa, M. Loewe, F. M\'endez and J. C. Rojas,
hep-th/0106125; J. Gamboa, M. Loewe and J. C. Rojas, {\it Phys.
Rev.} {\bf D64}, 067901 (2001); J. Gamboa, M. Loewe, F. M\'endez
and J. C. Rojas, {\it Mod. Phys. Lett.} {\bf 16A}, 2075 (2001); A.
P. Polychronakos, hep-th/0010264; S. Gubser and  M. Rangamani,
hep-th/0012155, V. P.  Nair and  A. P. Polychronakos, hep-th/0011172; D. Karabali, V. P.  Nair and  A. P. Polychronakos, {\it Nucl. Phys.}  {\bf B627} , 565(2002); O. F. Dayi,
L. T. Kelleyane,  hep-th/0202062; O. F. Dayi and  A.  Jellal,
hep-th/0111267; P. A. Horvathy, hep-th/0201007; D. H.  Correa, G.
S. Lozano , E.F. Moreno, F.A. Schaposnik,  hep-th/0105085;  H. R.
Christiansen, F. A. Schaposnik, hep-th/0106181; R. Iengo and  R.
Ramachandran , {\it JHEP}, 0202 (2002).
\bibitem{mol} C. A. Mead and D. G. Truhlar, {\it J. Chem. Phys.}
{\bf 70 (05)}, 2284 (1979); M. Berry, {\it Proc. Lond.}
{\bf A392}, 45 (1984).
\bibitem{fw} See {\it e.g.} A. Shapere and F. Wilczek,
{\it Geometric Phases in Phases}, World Scientific 1989.
\bibitem{ab} Y. Aharonov and D. Bohm, {\it Phys. Rev.} {\bf 115}, 485 (1958).
\bibitem{vitorio} J. Gamboa and V. O. Rivelles, {\it J. Phys.}
{\bf 24A}, L659 (1991).
 \bibitem{pesh} M. Peshkin, I. Talmi and L. J.  Tassie,
 {\it Ann. Phys.}  {\bf 16}, 426 (1961).
 \bibitem{aw} M. G. Alford and F. Wilczek, {\it Phys. Rev. Lett.}
 {\bf 62}, 1017 (1988).
\bibitem{carroll} S. Carroll, J. Harvey, V. A. Kostelecky,
C. D. Lane and T. Okamoto, {\it Phys. Rev. Lett.} {\bf 87},
141601 (2001).
\bibitem{chaichian} M. Cheichian, A. Demishev, P. Presnajder,
M. Cheikh-Jabbari and A. Tureanu, hep-th/
0012175.
\bibitem{pis}H. Falomir and P. A. G. Pisani,  {\it Jour. Phys. A }: Math Gen. 34, 4143 (2001).
\bibitem{lajacky} P. de S.  Gerbert and  R. Jackiw  {\it Commun. Math. Phys.}{\bf 124}, 229 (1989). 
\bibitem{ruso}A. Yelnikov, hep-th/0112134. 
\bibitem{taylor} See {\it e.g.} J. Taylor, {\it Scattering Theory},
J. Wiley and Sons (1972).
\bibitem{rui} S. N. M. Ruijsenaars, {\it Ann. Phys} (N.Y.) {\bf 146}, 1 (1983).
\bibitem{jackiw} See {\it e.g.} S. Deser and R. Jackiw, {\it Comm. Math. Phys.}
{\bf 118}, 495 (1988).
\bibitem{hagen}C.R. Hagen, {\it Phys. Rev. }{\bf D41}, 2015 (1990).
\bibitem{soldati}P. Giacconi, F. Maltoni, R. Soldati,{\it Phys. Rev.}{\bf D53}, 952 (1996).
\bibitem{conf} E. L. Feinberg, {\it Soviet Phys. Usp.} {\bf 5},
753 (1963); E. Corinaldesi and F. Rafeli, {\it Am.
J. Phys.} {\bf 46}, 1255 (1978).
\bibitem{glr}  J. Gamboa, M. Loewe and J. C. Rojas, hep-th/0101081.
\bibitem{fw2} See {\it e.g.} F. Wilczek, {\it Fractional Statistics
and Anyon Superconductivity}, World Scientific (1989).
\bibitem{tono} A. Tonomura, {\it Phys. Rev. Lett.} {\bf 48}, 1443 (1983);
see also M. Peshkin and A. Tonomura,
 {\it The Quantum Hall Effect}, Springer Verlag (1989).
\bibitem{dewitt} M. G. Laidlaw and C. M. de Witt, {\it Phys. Rev. }
{\bf D3}, 1375 (1971); L. S. Schulman, {\it
Phys. Rev.}{\bf D176}, 1558 (1968).
\end{references}
\end{document}